\documentclass[acmsmall,screen,authordraft,nonacm,review=false,timestamp=false]{acmart}
\AtBeginDocument{%
  }

\setcopyright{acmlicensed}
\copyrightyear{2018}
\acmYear{2018}
\acmDOI{XXXXXXX.XXXXXXX}

\acmConference[Conference acronym 'XX]{Make sure to enter the correct
  conference title from your rights confirmation emai}{June 03--05,
  2018}{Woodstock, NY}
\acmISBN{978-1-4503-XXXX-X/18/06}

\usepackage{booktabs} 
\usepackage{multirow}
\usepackage{colortbl}
\definecolor{verylightgray}{rgb}{0.92,0.92,0.92}
\newcommand{\bc}[1]{\rowcolor{verylightgray}#1}
\newcommand{\dg}{$^\dagger$} 
\usepackage{amsmath}
\usepackage[capitalise]{cleveref}

\newcommand{\R}{\ensuremath{\mathbb{R}}}

\newcommand{\hide}[1]{\iffalse {#1} \fi}

\everypar\expandafter{\the\everypar\looseness -1}
\begin{document}

\title{Static Pruning in Dense Retrieval using Matrix Decomposition}


\author{Federico Siciliano}
\orcid{0000-0003-1339-6983}
\email{siciliano@diag.uniroma1.it}
\affiliation{%
  \institution{Sapienza University of Rome}
  \city{Rome}
  \country{Italy}
}

\author{Francesca Pezzuti}
\orcid{0009-0005-2364-2043}
\email{f.pezzuti@studenti.unipi.it }
\affiliation{%
  \institution{University of Pisa}
  \city{Pisa}
  \country{Italy}
}

\author{Nicola Tonellotto}
\orcid{0000-0002-7427-1001}
\email{nicola.tonellotto@unipi.it}
\affiliation{%
  \institution{University of Pisa}
  \city{Pisa}
  \country{Italy}
}

\author{Fabrizio Silvestri}
\orcid{0000-0001-7669-9055}
\email{fsilvestri@diag.uniroma1.it}
\affiliation{%
  \institution{Sapienza University of Rome}
  \city{Rome}
  \country{Italy}
}

\begin{abstract}
  In the era of dense retrieval, document indexing and retrieval is largely based on encoding models that transform text documents into embeddings. The efficiency of retrieval is directly proportional to the number of documents and the size of the embeddings. 
Recent studies have shown that it is possible to reduce embedding size without sacrificing -- and in some cases improving -- the retrieval effectiveness. However, the methods introduced by these studies are query-dependent, so they can't be applied offline and require additional computations during query processing, thus negatively impacting the retrieval efficiency.
In this paper, we present a novel static pruning method for reducing the dimensionality of embeddings using Principal Components Analysis. This approach is query-independent and can be executed offline, leading to a significant boost in dense retrieval efficiency with a negligible impact on the system effectiveness.
Our experiments show that our proposed method reduces the dimensionality of document representations by over $50$\% with up to a $5$\% reduction in NDCG@10, for different dense retrieval models.
\end{abstract}



\keywords{Dense Retrieval, Static Pruning, Matrix Decomposition}


\maketitle

\section{Introduction}
With the rise of dense retrieval methods \cite{zhao2024dense},fficient indexing and retrieval of documents from large corpora has become increasingly important \cite{arabzadeh2021predicting}. Current approaches rely on encoding models that transform text documents into high-dimensional embeddings for effective retrieval \cite{xiong2021ance,hofstatter2021tasb,izacard2022unsupervised}. However, the computational complexity of query processing and the space occupation of embedding indexes grow with the number of documents and the dimensions of the embeddings \cite{fang2024scaling}, becoming a major bottleneck, especially in real-time applications \cite{kim2022applications}.

Over the years, various techniques have been proposed to address this trade-off between efficiency and effectiveness in dense retrieval. Several studies have shown that it is possible to prune the overall space required to store document embeddings.
One possibility is to prune document embeddings completely. 
For example, \citet{acquavia2023static} achieved significant space savings by statically pruning embeddings related to low-IDF tokens, reducing the index size by $45$\% on the MSMARCO passage ranking task, while maintaining retrieval performance in terms of NDCG@10 and MAP. Similarly, \citet{chang2024neural} proposed a static pruning method that estimates passage quality, allowing up to $25$\% of passages in a corpus to be discarded without compromising retrieval results across multiple benchmarks.
Another possibility is to prune selected embedding dimensions. \citet{faggioli2024dimension} proposed dimension importance estimators to assess, at query processing time, the contribution of individual dimensions in high-dimensional embeddings to overall ranking quality. On a per-query basis, their method selects a subset of dimensions to retain, leading to a performance improvement of up to $11.5$\% on nDCG@10 compared to using all dimensions. 

Despite these advances, static pruning approaches~\cite{acquavia2023static, chang2024neural} focus on document pruning, and depend on the passages content and on token-level embeddings. The dynamic pruning approach~\cite{faggioli2024dimension} focuses on dimension pruning, but relies on query-specific information, requiring additional computation at inference time. 

Starting from the problems identified above, in this paper, we address the following question: ``\textit{Can dimension pruning be achieved without relying on query-specific information or adding computational overhead at inference time?}''. To address this, we propose a novel approach for dimensionality reduction in embeddings, utilizing Principal Components Analysis (PCA)~\cite{pearson1901principal} as a possible dimensionality reduction mechanism. Our approach is orthogonal w.r.t. content-dependent static document pruning and query-dependent dynamic dimension pruning. Also, being query-independent, it can be performed offline. This offers significant advantages in terms of reduced space occupation and query latency.
Despite Latent Semantic Analysis (LSA) \cite{dumais1988using} is related to both PCA and singular value decomposition, our approach uniquely targets dense embedding matrices rather than sparse term-document matrices as in LSA, focusing on latent vector spaces instead of text semantics.

Our extensive experiments on several dense retrievers and query sets show that our proposed PCA-based static pruning method can reduce the dimensionality of document representations by more than $50$\%, i.e., a space efficiency of $2\times$, with an almost negligible impact on retrieval quality, i.e., at most a $5\%$ reduction in nDCG@10 across all models and query sets.


\section{Methodology}
Let $q$ and $\{d_1, \ldots, d_n\}$ denote a query and a corpus of $n$ documents represented in the latent space $\R^d$ by the bi-encoder of a dense neural model.
For a given query $q$, a dense retriever ranks documents by decreasing relevance with respect to $q$. The relevance score for each document $d_i$ is calculated as the dot product between $q$ and $d_i$: $s(q,d_i)=q^\top d_i$. This score reflects how closely $d_i$ aligns with $q$ in the embedding space, with higher scores indicating greater relevance.

We organise document embeddings into a matrix $D \in \R^{n \times d}$, where each row corresponds to a document embedding. This matrix represents our \textit{embedding index}, that will be stored in a system implementing (approximate) nearest neighbour search, such as FAISS~\cite{faiss}.
Hence, the relevance scores of the whole corpus w.r.t. a query can be computed as $s(q) = Dq$, where $s(q) \in \R^{n}$ contains the dot products of the query embedding w.r.t. each document embedding. At inference time, the query processing time is composed by the dot product, with average time complexity $O(dn)$, and the score sorting, with average time complexity $O(n+k\ log\ k)$\footnote{Assuming the QuickSelSort\cite{martinez2004partial} algorithm is used to sort the top-$k$ documents.}. The space occupancy of embedding index is $O(dn)$ bytes.

To reduce the query processing time and/or the embedding index size, we propose an approach that reduces the dimensionality of $D$ by applying PCA. First, we compute the eigendecomposition of the covariance matrix of $D$, such that $D^TD = W\Lambda W^T$, where $W \in \R^{d \times d}$, $\Lambda \in \R^{d \times d}$. $\Lambda$ is a diagonal matrix with the key property of having its main diagonal elements, called eigenvalues, sorted in decreasing order, in such a way that the first columns of the matrix $T=DW$ explain more variance than the subsequent ones.

In fact, the matrix $T$ provides a new representation of the $n$ documents, where the new embeddings have dimensions sorted by \textit{importance}. We use the ordering induced by $\Lambda$ on the original $d$ dimensions to prune the least important dimensions, and keep just the $m$, with $m  < d$, first columns of $T$ in a matrix $T_m \in \R^{n \times m}$. 
In doing so, we obtain a new embedding matrix, i.e., an index, $\hat{D} = T_m \in \R^{n \times m}$, computed offline, whose rows correspond to new, pruned document embeddings at \textit{cutoff} $c=(d-m)/d$.


To use this reduced matrix $\hat{D}$ to process queries, we use the first $m$ columns of $W$, i.e., $W_m \in \R^{d \times m}$, to transform a query $q$ to $\hat{q} = W_m^T q \in \R^{m \times 1}$. 
In doing so, the relevance scores over the whole reduced matrix $\hat{D}$ w.r.t. a query $\hat{q}$ becomes $s(\hat{q}) = \hat{D}\hat{q} = (DW_m)(W_m^Tq)$. At inference time, the query processing now has an average time complexity of $O(dm + mn)$, due to query transformation and dot product computation over $\hat{D}$, and score sorting with unchanged $O(n+k\ log\ k)$ time complexity. The space occupancy of the embedding matrix is $O(mn)$ bytes, and the transformation matrix $W_m$ occupancy is $O(md)$ bytes. So, the overall space occupancy is $O(mn + md)$. Since $m < d \ll n$, the query processing speed-up is $O(d/m)$, and the space reduction is $O(m/d)$.

Using the transformation $V_m$, computed on a corpus $D$ and applied to the queries, we can transform and prune the embeddings of another corpus $D' \in \mathbb{R}^{n' \times d}$: $\hat{D'} = D' V_m \in \mathbb{R}^{n' \times m}$.  We refer to this process as \textit{out-of-domain PCA}, as it enables us to transfer the dimensionality reduction learned from one dataset to another, retaining relevant features with fewer dimensions.

\section{Experimental Setup}
We restate our central question: “Can dimension pruning be achieved without relying on query-specific information or adding computational overhead at inference time?” To address this, we break down the main question into the following research questions, each explored through targeted experiments:
\begin{description}
    \item[RQ1] To what extent can the proposed in-domain PCA static pruning be applied without compromising the ranking effectiveness of the system?
    \item[RQ2] What is the impact of the proposed out-of-domain PCA static pruning on ranking effectiveness?
    \item[RQ3] How does the number of embeddings used to compute the PCA transformation matrix, impact the ranking effectiveness at various pruning cutoffs?
\end{description}

    In our experiments, we use ANCE~\cite{xiong2021ance}, TAS-B~\cite{hofstatter2021tasb}, and Contriever~\cite{izacard2022unsupervised} as bi-encoder models for computing the embeddings, and we use the FAISS toolkit~\cite{faiss} for indexing these embeddings and rank them at query processing time.
    To evaluate the in-domain ranking effectiveness, we use the query sets TREC DL 19~\cite{craswell2020dl2019}, DL 20~\cite{craswell2021dl2020}, DL HARD~\cite{mackie2021dlhard}, and MS MARCO Dev Small~\cite{bajaj2016msmarco}, and we use the passage collection of MS MARCO~\cite{bajaj2016msmarco}. To evaluate the effectiveness in out-of-domain passage ranking tasks, we use the BEIR TREC COVID~\cite{thakur2021beir} query set and its passage collection.
    We use \textit{ir\_measures}~\cite{MacAvaney2022irmeasures} to measure the classical effectiveness metrics AP, nDCG@10, MRR@10; in the remainder of this work, when omitted, we will assume a cutoff @10.
    The baselines for our experiments are the ranking lists produced by dense bi-encoders with no PCA transformation and no pruning.
    For significance testing w.r.t. the baseline, we apply a two-tailed paired Wilcoxon signed-rank test with $\alpha = 0.05$.












\section{Experimental Results}
    \subsubsection*{RQ1. In-Domain PCA.}
        \begin{table*}[ht!]
\centering
\caption{Pruning of dimensions via PCA computed on $10^5$ in-domain embeddings. $\dagger$~denotes significant differences w.r.t. the baseline. \textbf{Bold} denotes the highest value.}
\label{tab:id_effectiveness_pca}
\resizebox{\textwidth}{!}{%
\begin{tabular}{@{}cccccccccccccccc@{}} 
\toprule
\multirow{2}{*}{$c~(\%)$} & \multicolumn{3}{c}{DL 19}   & \multicolumn{3}{c}{DL 20}   & \multicolumn{3}{c}{DL HARD} & \multicolumn{3}{c}{DEV SMALL}  & \multicolumn{3}{c}{COVID}    \\ 
\cmidrule(l){2-16}
                       & AP       & MRR   & nDCG     & AP       & MRR   & nDCG     & AP       & MRR   & nDCG     & AP       & MRR      & nDCG     & AP       & MRR   & nDCG      \\ 
\midrule
\multicolumn{16}{c}{TAS-B}  \\
\midrule
\bc –-                 & \textbf{.4059} & \textbf{.8851} & .7175          & \textbf{.4509} & .8342          & .6837          & .2207          & .5034          & .3758          & \textbf{.3533}    & \textbf{.3469} & \textbf{.4102} & .0460          & .4899          & .4382           \\
25                     & .4055          & \textbf{.8851} & \textbf{.7180} & .4508          & \textbf{.8427}          & \textbf{.6856} & .2217          & \textbf{.5071} & \textbf{.3768} & .3523             & .3457          & .4089          & \textbf{.0460} & \textbf{.4930}          & .4387           \\
50                     & .4033          & .8752          & .7101\dg       & .4477          & .8405          & .6779          & \textbf{.2226} & .4975          & .3702          & .3525\dg          & .3462          & .4090          & .0452          & .4915          & \textbf{.4398}           \\
75                     & .3478\dg       & .8439          & .6547\dg       & .4224\dg       & \textbf{.8528} & .6424          & .1912\dg       & .4920          & .3253\dg       & .3108\dg          & .3036\dg       & .3638\dg       & .0361\dg       & .4449          & .3781\dg        \\ 
\midrule
\multicolumn{16}{c}{Contriever}                                                                                                                                                                                                                                                           \\ 
\midrule
\bc –-                 & \textbf{.4019} & .8140          & \textbf{.6744} & .4483          & .7997          & \textbf{.6716} & \textbf{.2223} & .5054          & .3772          & \textbf{.3470}    & \textbf{.3406} & \textbf{.4070} & \textbf{.0481} & \textbf{.6404} & \textbf{.4963}  \\
25                     & .4000          & \textbf{.8178} & .6671          & \textbf{.4492} & \textbf{.8088} & .6679          & .2217          & \textbf{.5316} & \textbf{.3778} & .3424\dg          & .3361\dg       & .4019\dg       & .0459\dg       & .6337          & .4889           \\
50                     & .3851\dg       & .7926          & .6524          & .4319\dg       & .7832          & .6669          & .2177          & .5122          & .3711          & .3301\dg          & .3237\dg       & .3881\dg       & .0438\dg       & .6356          & .4844           \\
75                     & .3243\dg       & .7616          & .5947\dg       & .3880\dg       & .7281          & .6242\dg       & .1761\dg       & .4505          & .3212\dg       & .2859\dg          & .2785\dg       & .3365\dg       & .0347\dg       & .6097          & .4314\dg        \\ 
\midrule
\multicolumn{16}{c}{ANCE}                                                                                                                                                                                                                                                                 \\ 
\midrule
\bc –-                 & .3321          & .8209          & .6459          & .3948          & \textbf{.7858} & .6441          & \textbf{.1853} & \textbf{.4599} & \textbf{.3262} & .3353             & .3297          & \textbf{.3872} & \textbf{.0358} & \textbf{.5243} & \textbf{.3819}  \\
25                     & .3321          & .8209          & .6459          & .3948          & \textbf{.7858} & .6441          & \textbf{.1853} & .4597          & .3261          & .3353             & .3297          & \textbf{.3872} & .0357          & .5133          & .3800           \\
50                     & \textbf{.3335} & \textbf{.8212} & \textbf{.6489} & \textbf{.3950} & .7856          & \textbf{.6442} & .1851\dg       & .4597          & .3260          & \textbf{.3354}\dg & \textbf{.3298} & \textbf{.3872} & .0355\dg       & .5136          & .3777           \\
75                     & .3281          & .8209          & .6430          & .3922          & .7715          & .6381          & .1819          & .4561          & .3198          & .3351\dg          & .3294          & .3863          & .0344\dg       & .5014          & .3673\dg        \\
\bottomrule
\end{tabular}
}
        
\end{table*}


        \cref{tab:id_effectiveness_pca} shows how the three metrics, across five query sets and three bi-encoders, change with pruning at $25\%$, $50\%$ and $75\%$ cutoffs, applied after performing PCA on $10^5$ documents from their in-domain corpora.

        As the cutoff increases, there is a controlled degradation in performance. Remarkably, even when pruning $50\%$ of the dimensions, the best-performing bi-encoder, \mbox{TAS-B}, does not suffer a statistically significant loss in effectiveness, while for the other bi-encoders, the decrease in NDCG@10 remains below $4\%$.
        Furthermore, ANCE shows no statistically significant differences even at a 75\% cutoff, underscoring its robustness under aggressive pruning.
        

        \begin{figure*}[htbp!]
            \centering
            \includegraphics[width=1.0\linewidth]{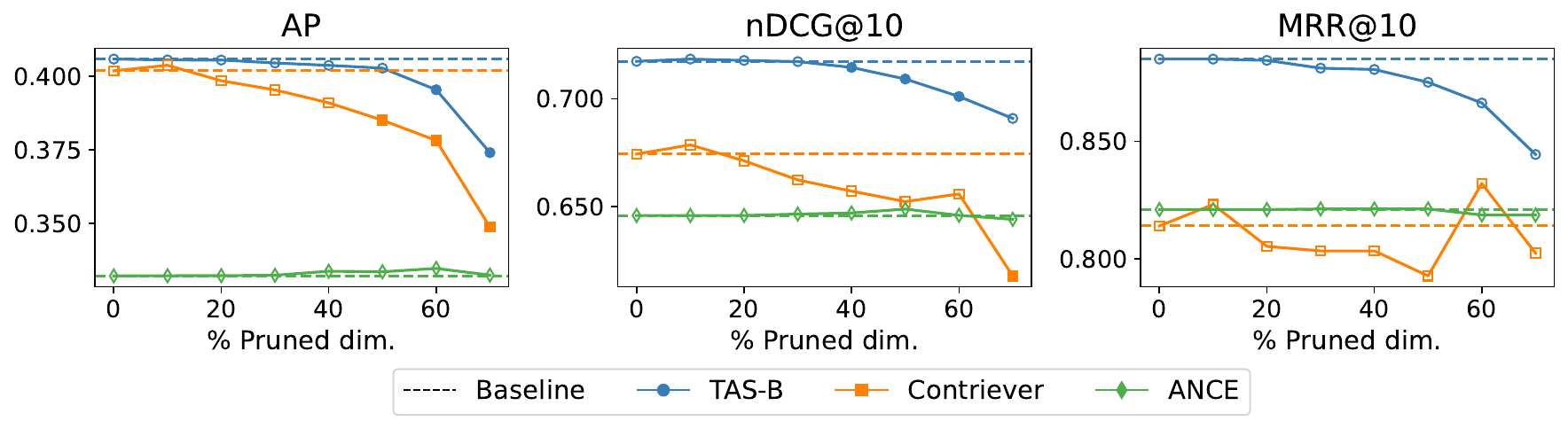}
            \caption{Effectiveness on DL 19 when applying PCA computed on $10^5$ in-domain embeddings and pruning dimensions at various cutoffs.
            \textit{Filled shapes} denote significant differences w.r.t. the baseline, whereas \textit{hollow shapes} represent non-significant differences. A significant difference followed by a non-significant one at a higher cutoff may result from increased variability as the metric mean decreases, with p-values often just above 0.05.}
            \label{fig:metrics-dl19-pca}
        \end{figure*}

        \cref{fig:metrics-dl19-pca} shows how three effectiveness metrics, measured on DL 19, vary when pruning the embeddings of the three bi-encoders at various cutoffs. Up to $50\%$ dimension pruning, none of the bi-encoders suffer from a degradation greater than $5\%$ in any of the metrics. Surprisingly, despite being the least effective model overall, ANCE retains much of its effectiveness across all metrics, even with a $70\%$ pruning. In contrast, Contriever seems to be the most sensitive bi-encoder to aggressive pruning. While we only present results for the DL 19 query set, similar trends were observed across all query sets.


    \subsubsection*{RQ2. Out-of-Domain PCA.}
        \begin{table*}[ht!]
\centering
\caption{Pruning of dimensions via PCA computed on $10^5$ out-of-domain embeddings. $\dagger$ denotes significant difference w.r.t. the baseline. \textbf{Bold} denotes the highest value.}
\label{tab:ood-effectiveness_pca}
\resizebox{\textwidth}{!}{%
\begin{tabular}{@{}cccccccccccccccc@{}} 
\toprule
\multirow{2}{*}{$c~(\%)$} & \multicolumn{3}{c}{DL 19}   & \multicolumn{3}{c}{DL 20}   & \multicolumn{3}{c}{DL HARD} & \multicolumn{3}{c}{DEV SMALL}  & \multicolumn{3}{c}{COVID}    \\ 
\cmidrule(l){2-16}
                       & AP       & MRR   & nDCG     & AP       & MRR   & nDCG     & AP       & MRR   & nDCG     & AP       & MRR      & nDCG     & AP       & MRR   & nDCG      \\ 
\midrule
\multicolumn{16}{c}{TAS-B}  \\
\midrule
\bc –-                 & \textbf{.4059} & \textbf{.8851} & .7175             & \textbf{.4509} & .8342          & \textbf{.6837} & .2207          & \textbf{.5034} & .3758          & \textbf{.3533}    & \textbf{.3469} & \textbf{.4102} & .0460          & .4899          & .4382              \\
25                     & .4045          & .8812          & \textbf{.7177}    & .4502          & \textbf{.8436} & .6829          & .2197          & .4985          & \textbf{.3770} & .3531             & .3465          & .4095          & \textbf{.0463} & .4931          & .4458\dg           \\
50                     & .4042          & .8812          & .7104             & .4489          & .8319          & .6781          & \textbf{.2234} & .4916          & .3711          & .3516\dg          & .3450\dg       & .4081\dg       & .0456          & .5022          & \textbf{.4467}\dg  \\
75                     & .3248\dg       & .8149          & .6142\dg          & .3786\dg       & .7667          & .6157\dg       & .1601\dg       & .4423          & .3013\dg       & .2878\dg          & .2802\dg       & .3378\dg       & .0384\dg       & \textbf{.5136} & .4105              \\ 
\midrule
\multicolumn{16}{c}{Contriever}                                                                                                                                                                                                                                                                 \\ 
\midrule
\bc –-                 & \textbf{.4019} & \textbf{.8140} & \textbf{.6744}    & \textbf{.4483} & .7997          & .6716          & \textbf{.2223} & .5054          & \textbf{.3772} & \textbf{.3470}    & \textbf{.3406} & \textbf{.4070} & \textbf{.0481} & \textbf{.6404} & \textbf{.4963}     \\
25                     & .3922\dg       & .8018          & .6679             & .4439          & \textbf{.8045} & \textbf{.6717} & .2147          & \textbf{.5105} & .3699          & .3401\dg          & .3335\dg       & .3988\dg       & .0471          & .6373          & .4897              \\
50                     & .3817\dg       & .8052          & .6572             & .4191\dg       & .7994          & .6511          & .1970\dg       & .4933          & .3558          & .3243\dg          & .3172\dg       & .3815\dg       & .0437          & .6242          & .4664              \\
75                     & .3374\dg       & .8082          & .6122\dg          & .3625\dg       & .7611          & .5752\dg       & .1761\dg       & .4550          & .3179\dg       & .2675\dg          & .2595\dg       & .3152\dg       & .0363\dg       & .5260          & .4158\dg           \\ 
\midrule
\multicolumn{16}{c}{ANCE}                                                                                                                                                                                                                                                                       \\ 
\midrule
\bc –-                 & .3321          & .8209          & .6459             & \textbf{.3948} & \textbf{.7858} & \textbf{.6441} & .1853          & .4599          & .3262          & .3353             & .3297          & .3872          & \textbf{.0358} & \textbf{.5243} & \textbf{.3819}     \\
25                     & .3322          & \textbf{.8212} & .6465             & \textbf{.3948} & \textbf{.7858} & \textbf{.6441} & \textbf{.1854} & \textbf{.4606} & \textbf{.3263} & .3354             & .3298          & .3873          & \textbf{.0358} & .5143          & .3815              \\
50                     & \textbf{.3338} & \textbf{.8212} & \textbf{.6500}\dg & .3947          & .7748          & .6427          & .1850          & .4570          & .3256          & .3359             & .3303          & \textbf{.3876} & .0357          & .5143          & .3806              \\
75                     & .3293\dg       & .8070          & .6394             & \textbf{.3948} & .7838          & .6428          & .1818          & .4444          & .3217          & \textbf{.3361}\dg & \textbf{.3305} & .3873          & .0354\dg       & .5031          & .3777              \\
\bottomrule
\end{tabular}
}
        
\end{table*}

        \cref{tab:ood-effectiveness_pca} presents the results for out-of-domain PCA, for which the embedding index is pruned using the transformation matrix computed on the embeddings of a different corpus. Surprisingly, the performance on all metrics is on par with that of the in-domain pruning. Even at a $50\%$ cutoff, TAS-B shows no significant degradation across most benchmarks, except on DEV SMALL, with differences still under $1\%$.

    
    \subsubsection*{RQ3. Number of embeddings used to compute PCA.}
    \begin{figure*}[!ht]
        \centering
        \includegraphics[width=1.0\linewidth]{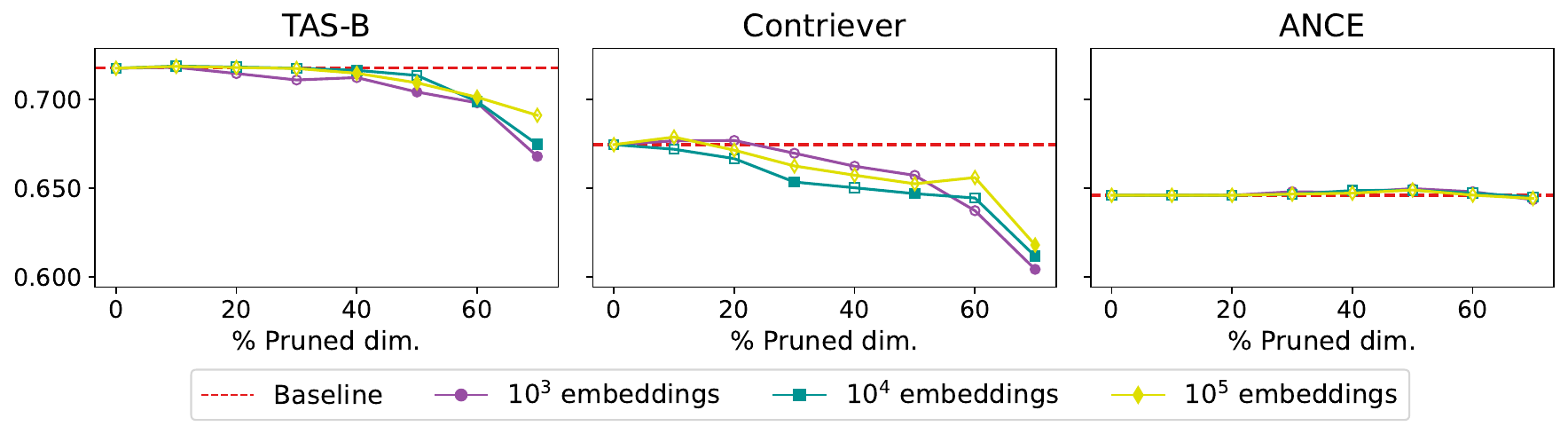}
        \caption{nDCG@10 on DL 19 when applying in-domain PCA computed on $10^3$, $10^4$ and $10^5$ embeddings, to prune embeddings at various cutoffs. \textit{Filled shapes} denote significant differences, whereas \textit{hollow shapes} represent non-significant differences. A significant difference followed by a non-significant one at a higher cutoff may result from increased variability as the metric mean decreases, with p-values often just above 0.05.}
        \label{fig:num_docs_ID_pca}
    \end{figure*}

    \cref{fig:num_docs_ID_pca}, for the three bi-encoders, shows nDCG@10 on DL 19 when varying the number of embeddings used for decomposition.
    All encoders show remarkable stability, with almost no significant differences from the baseline, even when the decomposition is performed on only $10^3$ documents, as differences between using $10^3$ and $10^5$ documents are negligible. While we only present results for the DL 19 query set, similar trends were observed across all query sets.

\section{Conclusions}
In study, we introduced PCA-based static dimension pruning for embedding indexes in dense retrieval systems to address our initial question ``\textit{Can dimension pruning be achieved without relying on query-specific information or incurring additional computational overheads at inference time?}''. The experimental results show that with our approach we can prune up to $50\%$ of embedding dimensions while minimally impacting retrieval quality. TAS-B, the most effective encoder taken into account, retains statistically unchanged performance after a $50\%$ reduction.
Furthermore, the effectiveness of our proposed method is not impacted by the number of documents used for decomposition, underlining its flexibility.

A key advantages of our approach is that it enables dimensionality reduction to be performed offline, allowing significant real-time efficiency gains during the inference phase without requiring additional computation per query. Furthermore, our method proves to be robust across different query sets and benchmarks, including out-of-domain corpora, demonstrating its generalisability.

Overall, this study highlights the potential of PCA as a powerful tool for optimising dense retrieval systems. The ability to significantly reduce both storage and computational costs while maintaining high retrieval quality suggests that PCA-based pruning can be a valuable addition to current retrieval pipelines.


\bibliographystyle{ACM-Reference-Format}

\bibliography{bibliography}


\begin{thebibliography}{20}


\ifx \showCODEN    \undefined \def \showCODEN     #1{\unskip}     \fi
\ifx \showDOI      \undefined \def \showDOI       #1{#1}\fi
\ifx \showISBNx    \undefined \def \showISBNx     #1{\unskip}     \fi
\ifx \showISBNxiii \undefined \def \showISBNxiii  #1{\unskip}     \fi
\ifx \showISSN     \undefined \def \showISSN      #1{\unskip}     \fi
\ifx \showLCCN     \undefined \def \showLCCN      #1{\unskip}     \fi
\ifx \shownote     \undefined \def \shownote      #1{#1}          \fi
\ifx \showarticletitle \undefined \def \showarticletitle #1{#1}   \fi
\ifx \showURL      \undefined \def \showURL       {\relax}        \fi
\providecommand\bibfield[2]{#2}
\providecommand\bibinfo[2]{#2}
\providecommand\natexlab[1]{#1}
\providecommand\showeprint[2][]{arXiv:#2}

\bibitem[Acquavia et~al\mbox{.}(2023)]%
        {acquavia2023static}
\bibfield{author}{\bibinfo{person}{Antonio Acquavia}, \bibinfo{person}{Craig Macdonald}, {and} \bibinfo{person}{Nicola Tonellotto}.} \bibinfo{year}{2023}\natexlab{}.
\newblock \showarticletitle{{Static Pruning for Multi-Representation Dense Retrieval}}. In \bibinfo{booktitle}{\emph{Proc. Eng}}. \bibinfo{pages}{1--10}.
\newblock


\bibitem[Arabzadeh et~al\mbox{.}(2021)]%
        {arabzadeh2021predicting}
\bibfield{author}{\bibinfo{person}{Negar Arabzadeh}, \bibinfo{person}{Xinyi Yan}, {and} \bibinfo{person}{Charles~LA. Clarke}.} \bibinfo{year}{2021}\natexlab{}.
\newblock \showarticletitle{{Predicting efficiency/effectiveness trade-offs for dense vs. sparse retrieval strategy selection}}. In \bibinfo{booktitle}{\emph{Proc. CIKM}}. \bibinfo{pages}{2862--2866}.
\newblock


\bibitem[Bajaj et~al\mbox{.}(2016)]%
        {bajaj2016msmarco}
\bibfield{author}{\bibinfo{person}{Payal Bajaj}, \bibinfo{person}{Daniel Campos}, \bibinfo{person}{Nick Craswell}, \bibinfo{person}{Li Deng}, \bibinfo{person}{Jianfeng Gao}, \bibinfo{person}{Xiaodong Liu}, \bibinfo{person}{Rangan Majumder}, \bibinfo{person}{Andrew McNamara}, \bibinfo{person}{Bhaskar Mitra}, \bibinfo{person}{Tri Nguyen}, \bibinfo{person}{Mir Rosenberg}, \bibinfo{person}{Xia Song}, \bibinfo{person}{Alina Stoica}, \bibinfo{person}{Saurabh Tiwary}, {and} \bibinfo{person}{Wang Tong}.} \bibinfo{year}{2016}\natexlab{}.
\newblock \showarticletitle{{MS MARCO: A Human Generated MAchine Reading COmprehension Dataset}}. In \bibinfo{booktitle}{\emph{InCoCo@NIPS}}.
\newblock


\bibitem[Chang et~al\mbox{.}(2024)]%
        {chang2024neural}
\bibfield{author}{\bibinfo{person}{Xuejun Chang}, \bibinfo{person}{Debabrata Mishra}, \bibinfo{person}{Craig Macdonald}, {and} \bibinfo{person}{Sean MacAvaney}.} \bibinfo{year}{2024}\natexlab{}.
\newblock \showarticletitle{{Neural Passage Quality Estimation for Static Pruning}}. In \bibinfo{booktitle}{\emph{Proc. SIGIR}}. \bibinfo{pages}{174--185}.
\newblock


\bibitem[Craswell et~al\mbox{.}(2021)]%
        {craswell2021dl2020}
\bibfield{author}{\bibinfo{person}{Nick Craswell}, \bibinfo{person}{Bhaskar Mitra}, \bibinfo{person}{Emine Yilmaz}, {and} \bibinfo{person}{Daniel Campos}.} \bibinfo{year}{2021}\natexlab{}.
\newblock \showarticletitle{{Overview of the TREC 2020 deep learning track}}. In \bibinfo{booktitle}{\emph{Proc. TREC}}.
\newblock


\bibitem[Craswell et~al\mbox{.}(2020)]%
        {craswell2020dl2019}
\bibfield{author}{\bibinfo{person}{Nick Craswell}, \bibinfo{person}{Bhaskar Mitra}, \bibinfo{person}{Emine Yilmaz}, \bibinfo{person}{Daniel Campos}, {and} \bibinfo{person}{Ellen~M. Voorhees}.} \bibinfo{year}{2020}\natexlab{}.
\newblock \showarticletitle{{Overview of the TREC 2019 deep learning track}}. In \bibinfo{booktitle}{\emph{Proc. TREC}}.
\newblock


\bibitem[Dumais et~al\mbox{.}(1988)]%
        {dumais1988using}
\bibfield{author}{\bibinfo{person}{Susan~T. Dumais}, \bibinfo{person}{George~W. Furnas}, \bibinfo{person}{Thomas~K. Landauer}, \bibinfo{person}{Scott Deerwester}, {and} \bibinfo{person}{Richard Harshman}.} \bibinfo{year}{1988}\natexlab{}.
\newblock \showarticletitle{{Using latent semantic analysis to improve access to textual information}}. In \bibinfo{booktitle}{\emph{Proc. SIGCHI}}. \bibinfo{pages}{281--285}.
\newblock


\bibitem[Faggioli et~al\mbox{.}(2024)]%
        {faggioli2024dimension}
\bibfield{author}{\bibinfo{person}{Guglielmo Faggioli}, \bibinfo{person}{Nicola Ferro}, \bibinfo{person}{Raffaele Perego}, {and} \bibinfo{person}{Nicola Tonellotto}.} \bibinfo{year}{2024}\natexlab{}.
\newblock \showarticletitle{{Dimension importance estimation for dense information retrieval}}. In \bibinfo{booktitle}{\emph{Proc. SIGIR}}. \bibinfo{pages}{1318--1328}.
\newblock


\bibitem[Fang et~al\mbox{.}(2024)]%
        {fang2024scaling}
\bibfield{author}{\bibinfo{person}{Yan Fang}, \bibinfo{person}{Jingtao Zhan}, \bibinfo{person}{Qingyao Ai}, \bibinfo{person}{Jiaxin Mao}, \bibinfo{person}{Weihang Su}, \bibinfo{person}{Jia Chen}, {and} \bibinfo{person}{Yiqun Liu}.} \bibinfo{year}{2024}\natexlab{}.
\newblock \showarticletitle{{Scaling laws for dense retrieval}}. In \bibinfo{booktitle}{\emph{Proc. SIGIR}}. \bibinfo{pages}{1339--1349}.
\newblock


\bibitem[Hofst{\"a}tter et~al\mbox{.}(2021)]%
        {hofstatter2021tasb}
\bibfield{author}{\bibinfo{person}{Sebastian Hofst{\"a}tter}, \bibinfo{person}{Sheng-Chieh Lin}, \bibinfo{person}{Jheng-Hong Yang}, \bibinfo{person}{Jimmy~J. Lin}, {and} \bibinfo{person}{Allan Hanbury}.} \bibinfo{year}{2021}\natexlab{}.
\newblock \showarticletitle{{Efficiently Teaching an Effective Dense Retriever with Balanced Topic Aware Sampling}}. In \bibinfo{booktitle}{\emph{Proc. SIGIR}}. \bibinfo{pages}{113--122}.
\newblock


\bibitem[Izacard et~al\mbox{.}(2022)]%
        {izacard2022unsupervised}
\bibfield{author}{\bibinfo{person}{Gautier Izacard}, \bibinfo{person}{Mathilde Caron}, \bibinfo{person}{Lucas Hosseini}, \bibinfo{person}{Sebastian Riedel}, \bibinfo{person}{Piotr Bojanowski}, \bibinfo{person}{Armand Joulin}, {and} \bibinfo{person}{Edouard Grave}.} \bibinfo{year}{2022}\natexlab{}.
\newblock \showarticletitle{{Unsupervised Dense Information Retrieval with Contrastive Learning}}.
\newblock \bibinfo{journal}{\emph{Proc. TMLR}} (\bibinfo{year}{2022}).
\newblock


\bibitem[Johnson et~al\mbox{.}(2021)]%
        {faiss}
\bibfield{author}{\bibinfo{person}{Jeff Johnson}, \bibinfo{person}{Matthijs Douze}, {and} \bibinfo{person}{Herv{\'e} J{\'e}gou}.} \bibinfo{year}{2021}\natexlab{}.
\newblock \showarticletitle{{Billion-{{Scale Similarity Search}} with {{GPUs}}}}.
\newblock \bibinfo{journal}{\emph{Trans. Big Data}} \bibinfo{volume}{7}, \bibinfo{number}{3} (\bibinfo{year}{2021}), \bibinfo{pages}{535--547}.
\newblock


\bibitem[Kim(2022)]%
        {kim2022applications}
\bibfield{author}{\bibinfo{person}{Yubin Kim}.} \bibinfo{year}{2022}\natexlab{}.
\newblock \showarticletitle{{Applications and future of dense retrieval in industry}}. In \bibinfo{booktitle}{\emph{Proc. SIGIR}}. \bibinfo{pages}{3373--3374}.
\newblock


\bibitem[{MacAvaney, Sean and Macdonald, Craig and Ounis, Iadh}(2022)]%
        {MacAvaney2022irmeasures}
\bibfield{author}{\bibinfo{person}{{MacAvaney, Sean and Macdonald, Craig and Ounis, Iadh}}.} \bibinfo{year}{2022}\natexlab{}.
\newblock \showarticletitle{Streamlining Evaluation with ir-measures}. In \bibinfo{booktitle}{\emph{Proc. ECIR}}. \bibinfo{pages}{305--310}.
\newblock


\bibitem[Mackie et~al\mbox{.}(2021)]%
        {mackie2021dlhard}
\bibfield{author}{\bibinfo{person}{Iain Mackie}, \bibinfo{person}{Jeffery Dalton}, {and} \bibinfo{person}{Andrew Yates}.} \bibinfo{year}{2021}\natexlab{}.
\newblock \showarticletitle{{How Deep is your Learning: the DL-HARD Annotated Deep Learning Dataset}}. In \bibinfo{booktitle}{\emph{Proc. TREC}}.
\newblock


\bibitem[Mart{\i}nez(2004)]%
        {martinez2004partial}
\bibfield{author}{\bibinfo{person}{Conrado Mart{\i}nez}.} \bibinfo{year}{2004}\natexlab{}.
\newblock \showarticletitle{Partial quicksort}. In \bibinfo{booktitle}{\emph{Proc. 6th ACMSIAM Workshop on Algorithm Engineering and Experiments and 1st ACM-SIAM Workshop on Analytic Algorithmics and Combinatorics}}. \bibinfo{pages}{224--228}.
\newblock


\bibitem[Pearson(1901)]%
        {pearson1901principal}
\bibfield{author}{\bibinfo{person}{Karl Pearson}.} \bibinfo{year}{1901}\natexlab{}.
\newblock \showarticletitle{{Principal components analysis}}.
\newblock \bibinfo{journal}{\emph{The Philosophical Magazine and Journal of Science}} \bibinfo{volume}{6}, \bibinfo{number}{2} (\bibinfo{year}{1901}), \bibinfo{pages}{559}.
\newblock


\bibitem[Thakur et~al\mbox{.}(2021)]%
        {thakur2021beir}
\bibfield{author}{\bibinfo{person}{Nandan Thakur}, \bibinfo{person}{Nils Reimers}, \bibinfo{person}{Andreas R{\"u}ckl{\'e}}, \bibinfo{person}{Abhishek Srivastava}, {and} \bibinfo{person}{Iryna Gurevych}.} \bibinfo{year}{2021}\natexlab{}.
\newblock \showarticletitle{{{BEIR}: A Heterogeneous Benchmark for Zero-shot Evaluation of Information Retrieval Models}}. In \bibinfo{booktitle}{\emph{Proc. NeurIPS nd Benchmarks Track}}.
\newblock


\bibitem[Xiong et~al\mbox{.}(2021)]%
        {xiong2021ance}
\bibfield{author}{\bibinfo{person}{Lee Xiong}, \bibinfo{person}{Chenyan Xiong}, \bibinfo{person}{Ye Li}, \bibinfo{person}{Kwok-Fung Tang}, \bibinfo{person}{Jialin Liu}, \bibinfo{person}{Paul~N. Bennett}, \bibinfo{person}{Junaid Ahmed}, {and} \bibinfo{person}{Arnold Overwijk}.} \bibinfo{year}{2021}\natexlab{}.
\newblock \showarticletitle{{Approximate Nearest Neighbor Negative Contrastive Learning for Dense Text Retrieval}}. In \bibinfo{booktitle}{\emph{Proc. ICLR}}.
\newblock


\bibitem[Zhao et~al\mbox{.}(2024)]%
        {zhao2024dense}
\bibfield{author}{\bibinfo{person}{Wayne~Xin Zhao}, \bibinfo{person}{Jing Liu}, \bibinfo{person}{Ruiyang Ren}, {and} \bibinfo{person}{Ji-Rong Wen}.} \bibinfo{year}{2024}\natexlab{}.
\newblock \showarticletitle{{Dense text retrieval based on pretrained language models: A survey}}.
\newblock \bibinfo{journal}{\emph{Trans. TOIS}} \bibinfo{volume}{42}, \bibinfo{number}{4} (\bibinfo{year}{2024}), \bibinfo{pages}{1--60}.
\newblock


\end{thebibliography}


\end{document}